\documentclass[prd, twocolumn, superscriptaddress, nofootinbib]{revtex4}
\usepackage{graphicx, epsfig}

\newcommand {\ds}{\displaystyle}

\def\fun#1#2{\lower3.6pt\vbox{\baselineskip0pt\lineskip.9pt
  \ialign{$\mathsurround=0pt#1\hfil##\hfil$\crcr#2\crcr\sim\crcr}}}

\begin{document}

\title{Is the Universe Inflating? Dark Energy and the Future of the Universe}

\author{Dragan Huterer}
\affiliation{Department of Physics,
Case Western Reserve University,
Cleveland, OH~~44106}
\author{Glenn D. Starkman}
\affiliation{Department of Physics,
Case Western Reserve University,
Cleveland, OH~~44106}
\author{Mark Trodden}
\affiliation{Department of Physics,
Syracuse University,
Syracuse, NY~~13244-1130}

\begin{abstract}
We consider the fate of the observable universe in the light of the
discovery of a dark energy component to the cosmic energy budget. We
extend results for a cosmological constant to a general dark energy
component and examine the constraints on phenomena that may prevent
the eternal acceleration of our patch of the universe. We find that
the period of accelerated cosmic expansion has not lasted long enough
for observations to confirm that we are undergoing inflation; such an
observation will be possible when the dark energy density has risen to
between 90\% and 95\% of the critical. The best we can do is make
cosmological observations in order to verify the continued presence of
dark energy to some high redshift. Having done that, the only
possibility that could spoil the conclusion that we are inflating
would be the existence of a disturbance (the surface of a true vacuum
bubble, for example) that is moving toward us with sufficiently high
velocity, but is too far away to be currently observable.  Such a
disturbance would have to move toward us with speed greater than about
$0.8c$ in order to spoil the late-time inflation of our patch of the
universe and yet avoid being detectable.
\end{abstract}

\preprint{SU-GP-02/2-1}

\maketitle

\section{Introduction}

There is now considerable evidence that the universe is dominated by a
peculiar energy component with negative pressure. This component,
called dark energy, leads to the acceleration of the universe and
explains why type Ia supernovae of intermediate redshift are observed
to be dimmer than they would be in a matter-only
universe~\cite{Riess:1998cb,Perlmutter:1998np}. Dark energy also
obviates the apparent discrepancy between large-scale structure
measurements, which indicate that matter comprises around 30\% of the
critical energy density, and cosmic microwave background measurements,
which show that the total energy density is very nearly equal to
critical. The energy density of this mysterious component, $X$,
relative to the critical density is $\Omega_X\sim 0.7$ and the equation of
state is $-1\leq w\equiv p_X/\rho_X \leq
-0.6$~\cite{Perlmutter:1999jt,Wang:1999fa}.

This important discovery raises some interesting and fundamental
issues. Of particular interest to us is the possibility that the
universe may be entering a stage of
inflation~\cite{Guth:1980zm,Linde:1981mu,Albrecht:1982wi}, similar to
that thought to have occurred in the early universe. If this is the
case, we would like to know when the universe started or will start to
inflate, when we will be able to observe this inflation, and what
observational constraints, if any, exist that could reveal, even in
principle, whether the inflationary period will be prolonged or even
eternal. To address these questions, we are motivated by the exciting
prospects for constraining dark energy using cosmological probes. Type
Ia supernovae (SNe Ia) have been the most effective and direct probes
to date, and give strong evidence for the existence of the
negative-pressure component~\cite{Riess:1998cb,Perlmutter:1998np}.
Number counts of galaxies~\cite{newman} and galaxy
clusters~\cite{holder} are also very promising techniques, which are
sensitive to the growth of density perturbations. While weak
gravitational lensing~\cite{WL} and large-scale structure
surveys~\cite{GDM} are mostly sensitive to the matter component, and
the cosmic microwave background (CMB) primarily probes the total
energy density, all three of these provide crucial complementary
information; namely the fraction of the total energy density in matter
$\Omega_M$ and the total energy density $\Omega_{TOT}$ (both in units
of the critical density).  With the proposed wide-field telescopes,
such as the Large-aperture Synoptic Survey Telescope
(LSST)\footnote{www.dmtelescope.org} on the ground, and the Supernova
Acceleration Probe (SNAP)\footnote{snap.lbl.gov} in space, the next
decade may offer an order-of-magnitude better constraints on the
properties of dark energy.

To address the observability of the fate of the universe, Starkman et
al.~\cite{Starkman:1999pg} (heretofore STV) have used the concept of
the minimal anti-trapped surface (MAS). The MAS is a sphere, centered
on the observer, on which the velocity of comoving objects is the
speed of light $c$. In a Friedmann-Robertson-Walker (FRW) cosmology,
the radius of the MAS at any given conformal time $\eta_e$ is the
Hubble radius at that time $H(\eta_e)^{-1}$.  For sources inside our
MAS, photons emitted directly at us get nearer with time, while all
photons emitted by sources outside the MAS are initially receding from
us because of the superluminal recession of the source.  If the MAS is
expanding (in comoving terms), then the retreating photons will
eventually stop retreating and reach the observer -- the source will
come into view.  If the MAS is contracting, many of the photons will
never reach us.  The authors then argue that comoving contraction of
the MAS can be identified with inflation.

The work of STV builds on earlier work of Vachaspati and
Trodden~\cite{Vachaspati:1998dy}, who have shown that in a FRW
cosmology at conformal time $\eta_e$, the necessary and sufficient
condition for the contraction of the MAS is that a region of size
$H^{-1}(\eta_e)$, the radius of the MAS, be vacuum dominated. Assuming
a flat universe with vacuum energy relative to critical of
$\Omega_{\Lambda}=0.8$, STV compute the redshift at which we can
observe the MAS to be $z_{MAS}\approx 1.8$. Since the comoving radius
of the MAS is equal to the comoving Hubble radius, the condition for
the onset of inflation is particularly simple
\begin{equation}
\frac{dH_{\rm comov}^{-1}}{d\eta}=0 \ .
\label{turnaroundcondition}
\end{equation}
STV then conclude that in a flat universe
the MAS is directly observable only if $\Omega_{\Lambda}>0.96$, and so,
for the currently favored value of $\Omega_{\Lambda}$, this is
not possible.

Subsequently several other paper on essentially the same topic have appeared,
and we briefly comment on them here. Avelino et al.\
\cite{Avelino:2000ix} address the same problem as STV, but replace the
criterion for inflation used by STV (i.e., the contraction of the MAS)
by the condition that inflation arises when the energy-momentum tensor
is vacuum-energy dominated out to a redshift $z=z_*$, where the
distance to $z_*$ is equal to the distance to the event horizon. With
this definition, higher $\Omega_{\Lambda}$ implies smaller $z_*$,
which the authors consider a more reasonable result than the one from
STV (in which higher $\Omega_{\Lambda}$ implies higher $z_{MAS}$.)
Gudmundsson and Bj\"{o}rnsson ~\cite{Gudmundsson:2001gd} introduce the
concept of a $\Lambda$-sphere, which they define as the surface within
which the vacuum energy dominates and is located at the redshift of
the onset of acceleration of the universe. In this work we shall
retain the criterion for inflation used by STV, defined by contraction
of the MAS. This definition is simple, mathematically precise and
intuitively clear and corresponds with the fundamental causal notion
of inflation -- that in inflation objects are leaving apparent causal
contact.

In this paper we extend the analysis of STV in several ways. First of
all, in Sec.~\ref{gencondition} we generalize all results from a
pure cosmological constant to a general dark energy component
described by its fractional density $\Omega_X$ and
redshift-dependent equation of state ratio $w(z)$. Indeed, although
the vacuum energy considered in STV is in many ways the simplest
dark-energy candidate, there are a number of other candidates, some of
which have been thoroughly explored
(e.g. quintessence~\cite{Ratra:1987rm,Caldwell:1997ii}, or
k-essence~\cite{Armendariz-Picon:2000dh,Armendariz-Picon:2000ah}). These
alternatives can have complex dynamics and lead to observationally
distinct cosmic evolutions.  In Sec.~\ref{toymodel} we use a toy
model to investigate in depth some possible scenarios. Finally, in
Sec.~\ref{observations} we discuss in detail what cosmological
observations can and cannot tell us about the fate of the
universe. Throughout, we assume a flat universe as suggested by recent CMB
anisotropy results~\cite{Netterfield:2001yq,Pryke:2001yz,max,wang}.
The fiducial model we use is $\Omega_X=1-\Omega_M=0.7$ and $w(z)=-1$,
which corresponds to the current concordance
model~\cite{turner_scripta}.

Finally, we make two additional assumptions. First, we assume the
validity of the weak energy condition (that is, $w(z)\ge -1$), which
is required in order to use the results of Vachaspati and Trodden
regarding the MAS~\cite{Vachaspati:1998dy}. Second, we assume that the
universe is homogeneous on the current horizon scales (i.e., on scale
$\sim H_0^{-1}$).  Besides being confirmed to a high accuracy by
observations, the homogeneity assumption is crucial for our arguments;
it follows that the energy content of our patch of the universe is a
function of time only (obtained through $\Omega_M(z)$ and
$\Omega_X(z)$) and not space.

\section{A Generalized condition for inflation}
\label{gencondition}
Let us see which dark energy models satisfy the
condition~(\ref{turnaroundcondition}) on the turnaround of the MAS,
and at what redshift.

The energy density in the dark component evolves as
\begin{equation}
\rho_X(z)=\Omega_X \rho_{crit}
\exp\left (3\int_0^z(1+w(z'))d\ln(1+z')\right ) \ .
\label{eq:rho_x}
\end{equation}
The condition for the turnaround of the MAS then simplifies to
\begin{equation}
\rho_M(z)+[1+3w(z)]\rho_X(z)=0,
\end{equation}

\noindent or, using the fact that
$\Omega_X(z)\equiv \rho_X(z)/\rho_{crit}(z)$,

\begin{equation}
Q(z)\equiv 1+3w(z)\Omega_X(z)=0 \ .
\label{eq:Q_condit}
\end{equation}

Since $dH_{comov}^{-1}/d\eta=0$ is equivalent to $\ddot a/a=0$ in an
FRW universe, this is precisely the same as the condition that the
universe be accelerating. Here $a(t)$ is the cosmic scale factor and a
dot denotes a derivative with respect to physical time $t$.

\begin{figure}
\epsfig{file=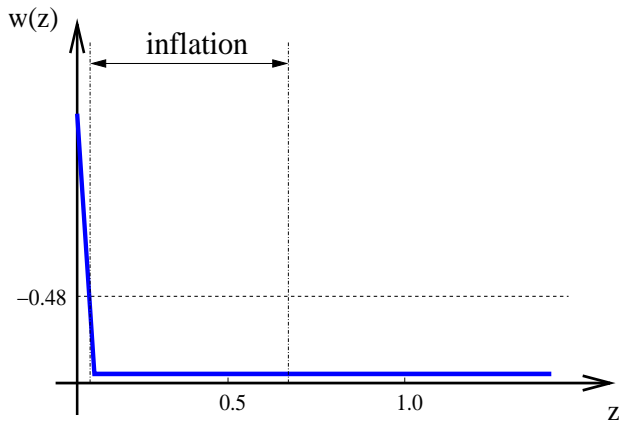, height=2.2in, width=3.2in}
\caption{Abrupt variations in $w(z)$ may stop (or commence)
inflation of our patch of the universe, yet be unobservable. An
example shown here is $w(z)$ that leads to inflation in the redshift
interval shown, but then suddenly increases. Such an
abrupt change in $w$ would be cosmologically unobservable. }
\label{fig:counterexmpl}
\end{figure}

Our observational knowledge of cosmic history puts certain constraints
on $Q(z)$ at large $z$.  It is possible that $\Omega_X(z)$ is
significant and $w(z)$ negative at particular eras at high $z$. For
example, during big bang nucleosynthesis $\Omega_X(z_{BBN})\lesssim
0.2$ is allowed \cite{ferreira}, and $\Omega_X(z)$ may even be unity
for short enough periods in a later era. However, there must have been
an epoch when $\Omega_X(z)$ was less than $1/3$ (the value we require
from Eq.~(\ref{eq:Q_condit})), since otherwise the dark energy
component would interfere with structure formation and big-bang
nucleosynthesis. Thus it is clear that $Q(z)>0$ at some large enough
$z$. In addition, the fact that galaxies and other objects in the
universe are visible tells us that the MAS was not contracting during the
same epoch. It then follows that the condition for inflation is
therefore $Q(z)<0$, or

\begin{equation}
w(z)<-{1\over 3\,\Omega_X(z)} \ .
\label{eq:w_condit}
\end{equation}
In particular, for $\Omega_X=0.7$ today, the criterion for inflation
is $w(0)<-0.48$, which coincides with the more familiar condition for
acceleration ($\ddot a/a>0$).

Of course, it is a significant experimental challenge to measure
$w(z)$ at a given $z$ or, more generally, to isolate the equation of
state ratio in a particular redshift window (for a more detailed
discussion, see Refs.~\cite{huttur,weller, kujat}).  Abrupt variations
in $w(z)$ are particularly difficult to detect due to the integral
effect of $w(z)$ on the expansion rate; see Eq.~(\ref{eq:rho_x}). An
example is given in Fig.~\ref{fig:counterexmpl}. The equation of state
ratio depicted here is negative at high $z$, causing the onset of
inflation (which occurred at $z=0.67$ if $w$ has always been $-1$);
$w$ then rapidly increases and inflation then stops when $w$ saturates
the bound of Eq.~(\ref{eq:w_condit}).  If the change in $w$ is
sufficiently abrupt, this change (and therefore the end of inflation)
will be cosmologically unobservable.

\section{A toy model}
\label{toymodel}

Our goal is to investigate a class of dark energy models to understand
in each case what observations may in principle tell us about the
future evolution of our patch of the universe. For example, since
current data indicate $\Omega_X\sim 0.7$ and $w\lesssim -0.8$ (see,
for example Ref.~\cite{bean}), implying current inflation, we would
like to understand how we might test the broad range of theoretical
scenarios that are consistent with current observations but in which
the MAS is {\it not} currently contracting.

\begin{figure}
\epsfig{file=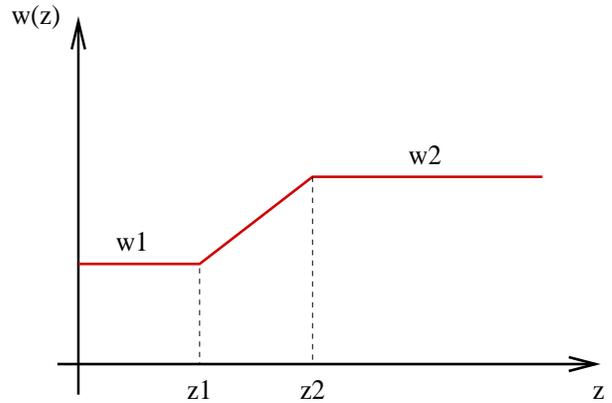, height=2.2in, width=3.2in}
\caption{The four-parameter ansatz for $w(z)$ used in this analysis.}
\label{fig:wz_model}
\end{figure}

We will adopt a four-parameter description of $w(z)$ of the following form
\begin{equation}
w(z)=\left\{\begin{array}{cl}
w_1 & \ \ \ \ \ 0<z<z_1 \\[0.1cm]
\ds{\frac{w_1(z_2-z) -w_2(z_1-z)}{z_2-z_1}}
                       & \ \ \ \ \ z_1\leq z \leq z_2 \\[0.3cm]
w_2 & \ \ \ \ \ z>z_2
\end{array} \right.
\label{eosansatz}
\end{equation}
(see Fig.~\ref{fig:wz_model}), so that the equation of state ratio
takes different constant values at low and high redshifts, and
interpolates linearly between these two regimes.  This
parameterization, although crude, mimics a wide class of dark energy
models. Although in principle we have four adjustable parameters
$w_1$, $w_2$, $z_1$ and $z_2$, the results in the $w_1$-$w_2$  plane are
similar for various choices of $z_1$ and $z_2$ and so it is not
necessary to run through the full parameter space.

\begin{figure}[!tb]
\epsfig{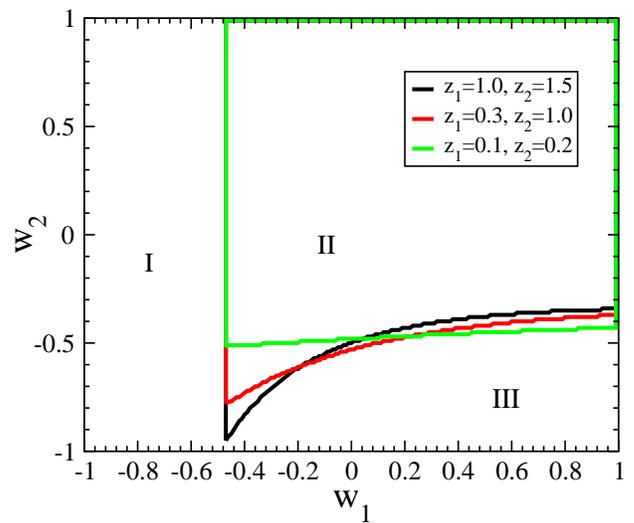}
\caption{The fate of the universe as a function of $w_1$ and $w_2$
in our toy model.  The results are fairly insensitive to $z_1$ and
$z_2$, and are plotted for three pairs of these parameters. Region I:
the universe is inflating now. Region II: the universe never
inflated. Region III: the universe inflated, but then stopped
inflating.}
\label{fig:no_turn}
\end{figure}

As Fig.~\ref{fig:no_turn} shows, there are three interesting
regions of parameter space.
\begin{itemize}
\item Region I: Here $w_1<-1/(3\Omega_X)$, and hence the universe is undergoing
inflation today independent of the behavior of $w(z)$ at higher $z$.
\item Region II: Here  $w_1$ is too high for inflation to be occurring today,
and $w_2$ is too high for inflation to have occurred in the past,
cf. Eq.~(\ref{eq:w_condit}). Hence, our patch of the universe never
underwent late-time inflation.
\item Region III: Here $w_1$ is too large for inflation to be
occurring now, but $w_2$ is negative enough that the universe recently
inflated but then stopped inflating.
\end{itemize}
Note also the $z_1$ and $z_2$-dependent ``dip'' at $w_1$ just greater
than $-0.48$. This feature can be explained simply. If $w_1=-0.45$,
say, then dark energy becomes subdominant to dark matter with
increasing redshift, and it takes a very negative $w_2$ (close to
$-1$) to achieve inflation in the past. However, if $w_1$ is more
positive, $+1.0$ say, dark energy is not as subdominant at higher $z$
(or perhaps is dominant), and it takes a less negative $w_2$ to
achieve inflation in the past.

\section{Observing the contraction of the MAS?}
\label{observations}

Let us begin by generalizing the condition of STV~\cite{Starkman:1999pg}
for the observability of the turnaround point of the MAS. The redshift
of the turnaround $z_c$ (where `c' stands for `contraction') is given by
\begin{equation}
Q(z_c)=0,
\label{eq:zc}
\end{equation}
where $Q(z)$ is given in Eq.~(\ref{eq:Q_condit}) and $\rho_X(z)$ is
given by Eq.~(\ref{eq:rho_x}). The condition for observing the turnaround point
is given by
\begin{equation}
a_c \int_{\eta_0}^{\eta_c} d\eta=H^{-1}(\eta_c) \ ,
\end{equation}
which, when combined with Eq.~(\ref{eq:zc}) and for constant $w$,
simplifies to
\begin{equation}
\int_{a_c}^1 {dx \over \sqrt{x^3(x^{3w}-1-3w)}}
=  {1\over \sqrt{-3w}} \ ,
\label{eq:see_MAS}
\end{equation}
where $a_c$ is the scale factor at turnaround (normalized to 1
today).

\begin{figure}[!ht]
\epsfig{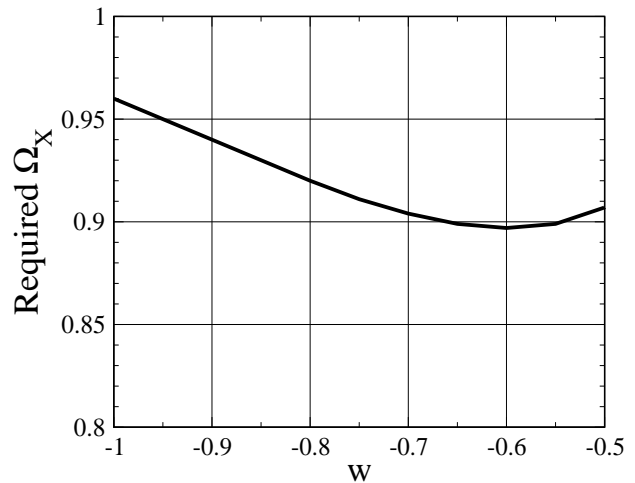}
\caption{The required minimum value of $\Omega_X$ in order for us
to be able to observe the turnaround point of the MAS today. Shown
as a function of (constant) $w$. }
\label{fig:required_OX}
\end{figure}

We can now compute which models allow the turnaround to be observable
(that is, $a_c<1$). Solving Eq.~(\ref{eq:see_MAS}) numerically for
$a_c$ and combining with Eq.~(\ref{eq:zc}), we find that, to a
good approximation, the constraint from STV is roughly independent of
w; see Fig.~\ref{fig:required_OX}.  Therefore, $\Omega_X\gtrsim 0.9-0.95$
would be necessary to directly observe the contraction of the MAS,
with weak dependence of this value on $w$. Sadly, current data suggest
that $\Omega_X\simeq 0.7$.

\begin{figure}[!ht]
\epsfig{file=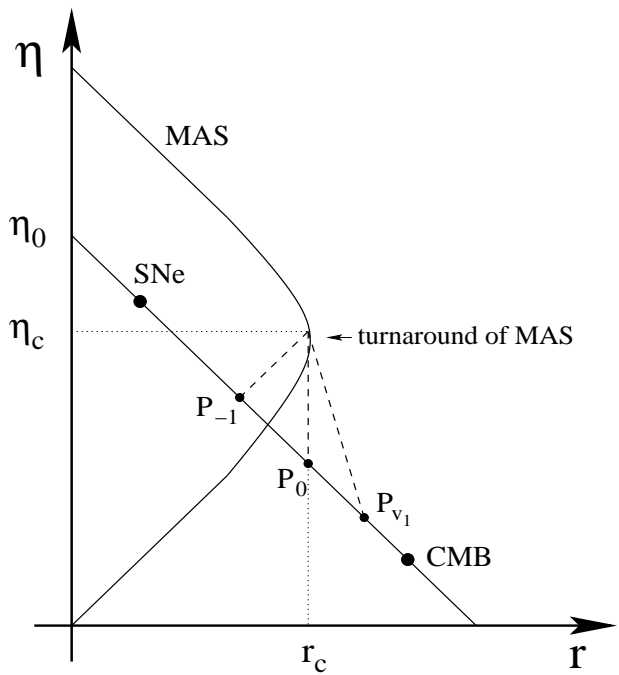, height=3.5in, width=3.2in}
\caption{Spacetime diagram showing the MAS position ($r=H_{comov}^{-1}$)
as a function of conformal time $\eta$. The universe undergoes
inflation at $\eta>\eta_c$.  On our present light-cone we observe
supernovae (SNe) and the cosmic microwave background (CMB), while we
presently can not observe the turnaround point of the MAS. The best we
can do is observe points on our light cone that are causally connected
with the turnaround point of the MAS. These points are denoted as
$P_v$, where $v$ is the speed of signal at $P_v$ required for it in
order to interact with the turnaround point. Three points are shown,
for $v=-1$ (signal moving away from as at the speed of light), $v=0$
(signal at rest in comoving coordinates) and $0<v=v_1<1$. }
\label{fig:MAS}
\end{figure}

Since the contraction of the MAS is currently unobservable, the best
we can hope for is to observe points on our light cone that are
causally connected with the turnaround point of the MAS. By making
observations out to one of these points, call it $P$, we can verify
(at least in principle) the continued presence of the dark-energy
component (i.e., $\Omega_X(z)$ and $w(z)$). Having successfully done
this, the only obstacle to concluding that the universe is inflating
is that there could exist some physics, such as a domain wall
separating our region of false vacuum from one of true vacuum, located
at $P$ and which adds enough energy (with sufficiently positive
pressure) at the turnaround point to spoil the contraction of the MAS,
and hence inflation.

The redshift of these observable points $P$ can be computed for a
given speed with which a signal located at that point (and observed by
us) moves in order to reach the turnaround of the MAS some time
later. In Fig.~\ref{fig:MAS}, these points are labeled as $P_v$, where
$v$ is the speed of the signal in question ($v$ is in units of the
speed of light $c$).

\begin{figure}[!ht]
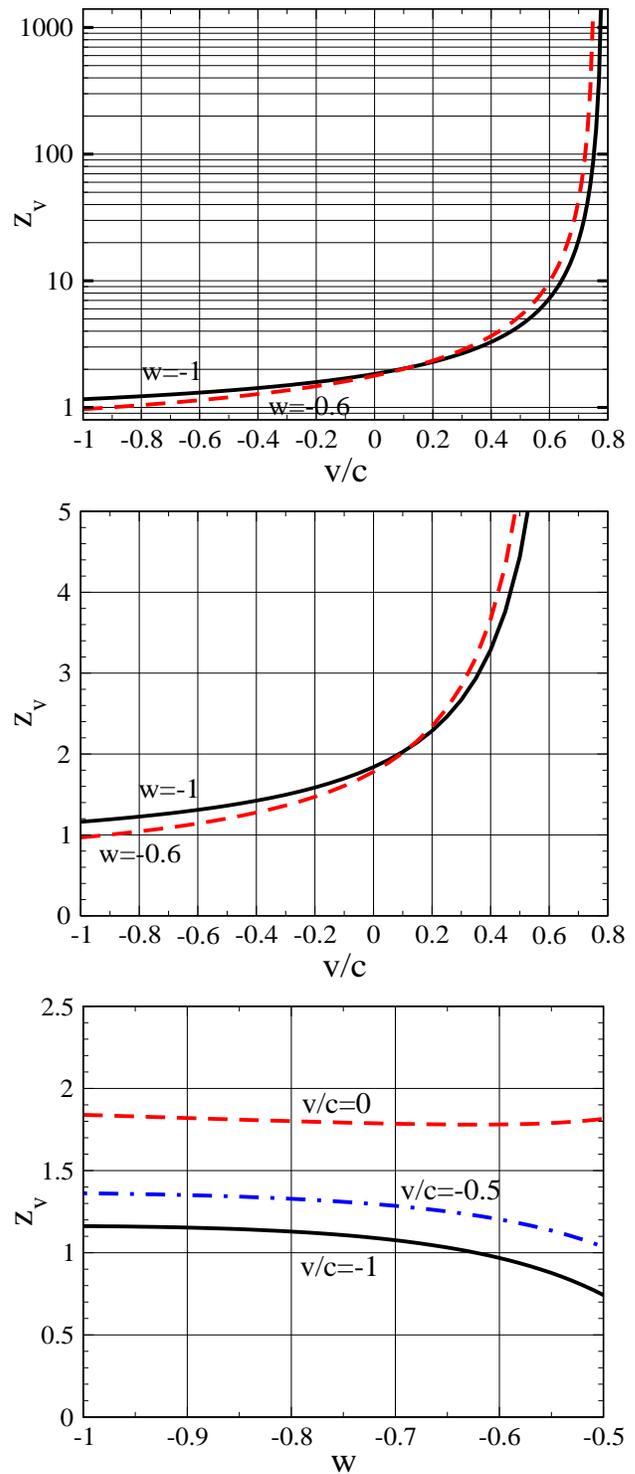

\epsfig{file=zv_vs_v.logplot.eps, height=2.5in, width=3.2in}\\[0.2cm]
\epsfig{file=zv_vs_v.eps, height=2.5in, width=3.2in}\\[0.2cm]
\epsfig{file=zv_vs_w.eps, height=2.5in, width=3.2in}
\caption{Top panel: $z_v$ computed as a function of $v$ for $w=-1$
and $w=-0.6$.  Middle panel: Same as top panel, but now shown with
low-$z_v$ region magnified. Bottom panel: $z_v$ computed as a function
of $w$ for $v=0$, $-0.5$ and $-1$.  All panels assume constant $w$.}
\label{fig:turnaround}
\end{figure}

From Fig.~\ref{fig:MAS}, the conformal distance and time at
which we see these points is given by
\begin{eqnarray}
r_v &=& {v(\eta_0-\eta_c)-r_c\over v-1}\label{eq:r_v} \\
\eta_v &=& \eta_0-r_v \ ,
\label{eq:eta_v}
\end{eqnarray}
where $v$ is the velocity of the signal with respect to us, $r$ and
$\eta$ are conformal distance and time respectively, and subscripts
$0$ and $c$ denote today and at the contraction point of the MAS
respectively. We have adopted the convention that $v=\pm 1$ denotes
the signal moving directly toward (away) away from us (at the speed
of light). Obviously, $r(v=1)=\infty$, since if $r(v=1)$ were finite,
then the MAS would be on our light cone and we would be able to see
it.

We solve Eqs.~(\ref{eq:r_v}) and (\ref{eq:eta_v}) numerically, and
display the results in Fig.~\ref{fig:turnaround}.  The top and middle
panels of Fig.~\ref{fig:turnaround} show the observable redshifts
$z_v$ as a function of the speed of the signal $v$ for two different
values of $w$. Note that $z_v$ increases rapidly as $v$ increases,
going to infinity for $v\sim 0.8$, and is fairly insensitive to $w$.
The bottom panel shows $z_v$ as a function of (constant) $w$ for three
different values of $v$; for example, for $v=0$ and $-1<w<-0.5$,
$z_v\approx 1.8$.

\begin{figure}[!ht]
\epsfig{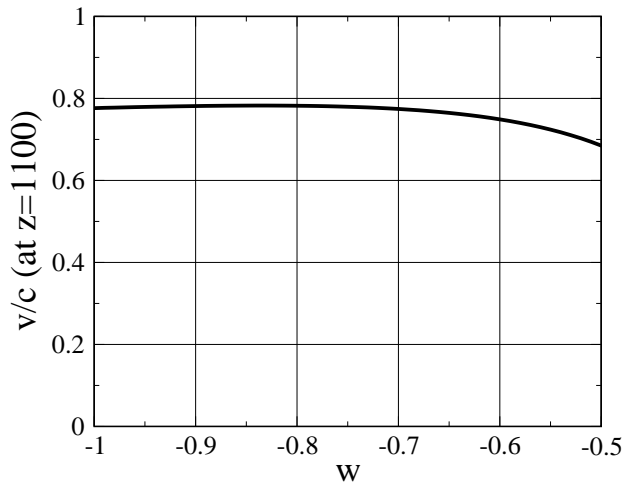}
\caption{The requisite velocity of a signal located at the last scattering surface
($z\approx 1100$) in order for it to interact with the turnaround
point of the MAS.}
\label{fig:v_z1100}
\end{figure}

Now, when we look at the CMB sky, we observe photons that arrive from
the last scattering surface (LSS). We compute the velocity with which
a signal located at the LSS would have to propagate in order for the
signal to interact with the turnaround point of the MAS, and hence
affect the onset of inflation in our patch of the universe. We use
Eqs.~(\ref{eq:r_v}) and (\ref{eq:eta_v}) to compute this velocity and
its dependence on various cosmological parameters. Representative
results are displayed in Fig.~(\ref{fig:v_z1100}) in which we plot $v$
as a function of constant $w$, for fixed $\Omega_X=0.7$. Clearly, for
any disturbance propagating at $v\gtrsim 0.8c$ it is impossible to
rule out that this disturbance has prevented or ended a recent
inflationary period.

\section{Conclusions}

During the last few years there has been considerable excitement over
a wide variety of data, most directly the observations of type Ia
supernovae, strongly suggesting that on the scales that have been
probed the rate of cosmic expansion is accelerating (or at least was
at $z\simeq 0.5$). If correct, this implies that the energy density of
the universe is (or was) dominated by dark energy -- a component with
negative pressure comparable to its energy density.  The inferences
that have been drawn -- that the entire universe is currently in the
throes of a new period of inflation of indefinite, and perhaps
infinite, duration -- rely on common but important simplifying
assumptions: that the dark energy density is homogeneous out to at
least the limit of the visible universe, and that the time evolution
of the dark energy density, if any, is relatively slow, governed
perhaps by the classical evolution of a scalar field in some smooth,
flat, effective potential.

Unfortunately, as we have found, none of these inferences or
assumptions are fully testable.  The period of accelerated cosmic
expansion has not lasted long enough for any observations, even in
principle, to confirm that the local Hubble volume is vacuum-dominated
and contained in the interior of an antitrapped surface -- the
condition that inflation is indeed taking place.  Such observations,
even assuming a homogeneous cosmological constant-dominated universe,
will need to wait until the energy density of the cosmic vacuum has
risen to about 95\% of the critical energy density from its current
70\%.  If the assumption of spatial homogeneity is maintained, but the
assumption of a static source of dark energy density is relaxed, then
we find that inflation is underway in any epoch in which $w$, the
effective equation of state of the dark energy over a Hubble volume, is
sufficiently vacuum-like; $w < -1/(3\Omega_x)$.  Investigations of
spatial inhomogeneities, particularly ones that could end any ongoing
inflationary expansion, require observations of relatively small
effects at relatively large distances.  It is possible, in principle,
to look out and tell whether a slow moving ($v\lesssim 0.8c$)
disturbance will prevent our little corner of the universe from
inflating. However, until $\Omega_X\simeq0.95$, we cannot be
completely confident that this inflation will ultimately begin and we
can never tell how long it will last, since this depends on the future
behavior of dark energy and its equation of state. The future of the
universe remains uncertain.

\begin{acknowledgments}
The authors would like to thank Tanmay Vachaspati for numerous helpful
discussions.  The work of DH and GDS is supported by a Department of
Energy grant to the particle astrophysics theory group at CWRU.  The
work of MT is supported by the National Science Foundation under grant
PHY-0094122.
\end{acknowledgments}

\end{document}